# Architected Porous Metals in Electrochemical Energy Storage

Vladimir Egorov[1] and Colm O'Dwyer[1,2,3,4]*

[1]School of Chemistry, University College Cork, Cork, T12 YN60, Ireland
[2] Micro-Nano Systems Centre, Tyndall National Institute, Lee Maltings, Cork, T12 R5CP, Ireland
[3]AMBER@CRANN, Trinity College Dublin, Dublin 2, Ireland
[4]Environmental Research Institute, University College Cork, Lee Road, Cork T23 XE10, Ireland

**Abstract**

Porous metallic structures are regularly used in electrochemical energy storage devices as supports, current collectors or active electrode materials. Bulk metal porosification, dealloying, welding or chemical synthesis routes involving crystal growth or self-assembly for example, can sometimes provide limited control of porous length scale, ordering, periodicity, reproducibility, porosity and surface area. Additive manufacturing and 3D printing has shown the potential to revolutionize the fabrication of architected metals many forms, allowing complex geometries not usually possible by traditional methods, but enabling complete design freedom of a porous metal based on the required physical or chemical property to be exploited. We discuss properties of porous metal structures in EES devices and provide some opinions on how architected metals may alleviate issues with electrochemically active porous metal current collectors, and provide opportunities for optimum design based on electrochemical characteristics required by batteries, supercapacitors or other electrochemical devices.

*Corresponding author: Email: c.odwyer@ucc.ie; Tel: +353 (0)21 4902732

**Keywords:** Porous Metals, Current Collectors, Additive Manufacturing (AM), Metal foams, Electrochemical Energy Storage (EES), Batteries, Supercapacitors, Metallic Lattice

# 1. Introduction

Porous metals, in ordered or random form, have been a mainstay in electrochemical science and technology[1-4]. Porous metals and metallic foam have been used as high surface area electrodes, current collectors, substrates, counter electrodes or even faraday cages. We now have a wide range of methods to fabricate porous metal with different porosity length scales, and with different degrees of ordering or periodicity[5-8]. While certain metals, their alloys and their physical properties fundamentally limit the range of latticed or porous structures that can be formed, the choice of metal for electrochemical energy storage applications[9-12] is usually based on surface activity, passivity, electrical conductivity and weight – the length scale of porosity[13], and the types of porous ordering are not always obtainable 'a la carte' for the electrochemical application when open-worked from the bulk metal. We examine the nature of porous metal formation, address some limitation for transition metals and noble metals in foam, lattice or porous form[14] for electrochemical energy storage devices and discuss how the choice of metal and the method of fabrication influence the nature of the porosity and their relative benefit in energy storage devices.

While directed self-assembly from chemical routes, sacrificial templates[15], dealloying approaches, or macropore formation methods can create a wide variety of porous structures, including metals, replication and minimization (or elimination) of structural defects is a persistent problem when they are used as electrodes[16, 17]. Additive manufacturing methods, including 3D printing, has evolved to allow large and small structures to be engineered to provide long range periodic or aperiodic metallic porous structures[18-21] that are, by comparison, almost defect free in their lattice structure. Accurate replication of any porous structure will allow for better comparison between supercapacitor, battery and other electrodes (e.g. water splitting and many more electrochemical processes[22]) to be quantitatively compared and assessed, provide a defined surface area for determining intrinsic geometrical surface area[23-25] effects in electrochemical processes, and allow us to choose metals and porous structures to take advantage of specific physics properties useful to the application, e.g. toughness, thermal conductivity, flexibility, surface activity etc.

Here, we provide a summary perspective on porous metal structures using in electrochemical energy storage (EES), particular supercapacitors and batteries, and the primary methods used to date to make these porous metals. We discuss periodic porous metals and porous metal foams, the main methods used to make these structures, and provide our opinion on the advent of additive manufacturing for metallic lattice or porous metal formation for EES devices. We also discuss the merits of some additive manufacturing approaches that can create metals in more complex and useful structures that cannot be achieved form porosification of bulk metals, dealloying and chemical assembly routes for example. These methods for porous metals widen the choice of active materials for electrochemical applications, or indeed as passive, conductive current collectors that do not contribute unwanted parasitic activity.

## 2. Porous Metal Fabrication Methods

Metallic porous materials can exist in a variety of forms, such as foams, sponges, micro/nanolattices, micro/nanoarrays etc.[19-21, 23, 24] Structurally, porous materials can be categorized as stochastic (random) and ordered (mostly periodic, but aperiodic 3D-tiling is also possible), depending on the randomness of pore size and arrangement (Fig. 1a) [5, 20]. Examples of stochastic structures include metal foams and sponges, whereas micro or nanolattices belong to the ordered category. In order to achieve high accessible specific surface area and create continuous diffusion pathways, which is important for EES applications, the porous structures should ideally be open-cell [5], i.e. internal pores should be well-interconnected minimizing the number of isolated cavities. Complete interconnectivity is more easily achieved in the engineered lattices and arrays particularly when structure is imposed by design. For other types of structures such as bottom-up- or self-assembled for example, interconnectivity is a function of the method and conditions of fabrication. Stochastic materials are usually isotropic at scales larger than a cell size, however, periodic structures can be highly anisotropic when the spatial arrangement or dimensions of unit blocks have a prevalent direction [20].

*Metallic foams and sponges* (open-cell foams) are traditionally produced using two main approaches: by creating pores in the bulk material or by fusing many separate cells into a single structure [26]. The former approach in turn can be implemented using many different techniques. *Gas injection* employs pumping of a pressurized gas into metal melt forming gas bubbles, which then stay after cooling and form pores. *Addition of foaming agents*, such as metal hydrides, carbonates, oxides, which release gases upon heating, is another technique for metal foam fabrication. *Space holder* and *template-directed* methods utilize various fillers (e.g. polymer beads, salt grains, hollow spheres, woven wire meshes) or templates (e.g. polymer foams, ceramic templates, etc.), which are subsequently removed, for example, by pyrolysis or solvent leaching. *Sponge replication method* can be viewed as a variation of the previous methods in which metallic powder slurry is impregnated into polymeric foam followed by pyrolysis of polymer and post-sintering [6]. *Dealloying*, a corrosion process in which one or several components of an alloy are dissolved, is a powerful method of fabrication of metal foams and sponges capable of producing micro-, nano- and hierarchical (micro-nano) pores [13, 27, 28].

*Periodic metallic structures* such as lattices or arrays can also be fabricated by several different methods. *Investment casting* is one of the traditional industrial techniques used in metallurgy capable of producing complex high-precision patterns [19, 29]. This method is based on creation of a sacrificial master pattern (e.g. by rapid prototyping methods) from a wax or polymer, which is then used to fabricate a mould by coating. Wax or polymer are subsequently removed by melting, pyrolysis or evaporation to leave a hollow mould, in which molten metal is later injected. *Deformation forming* can involve metal sheet perforation, deformation by pressing, shearing, expanding and layer stacking. The unit cell size achievable by this technique can be as low as millimeter scale, which is quite large for micro- or nanolattice fabrication. Using *metal textiles (woven or non-woven)* is another conventional approach to manufacturing periodic lattice materials, where metal wires are intertwined at various orientations, bent, sheared or otherwise deformed and, optionally, soldered [21]. These approaches are relatively fast and inexpensive, but in some cases are limited by the specifications and precision of the equipment and forming methods.

The traditional techniques for metallic lattice fabrication have a number of disadvantages, such as a relatively high cost and complexity, slow fabrication rate, low number of cells per unit length (hence, large feature size), waste of sacrificial materials, limited choice of suitable low-viscosity metals or alloys (in the case of investment casting). Due to the nature of fabrication process, some methods have limited capability for producing complex geometrical structures.

*Additive manufacturing (AM),* or *3D printing*, which is a common name for a family of manufacturing methods utilizing layer-by-layer fabrication of three-dimensional models created with computer aided design (CAD) [3], has been gaining popularity in the fabrication of complex shapes and objects, in particular periodic metallic structures. Contrary to conventional techniques we have just discussed, AM is highly versatile and enables fabrication of both stochastic and ordered open-cell structures with mono- or hierarchical scaling of pore sizes and these will prove useful for new form factor batteries and supercapacitors[30]. Given the ongoing improvements in 3D printing hardware, AM is able to produce high-resolution, geometrically accurate objects having small cell unit size at low cost and relatively fast production rate. This makes AM a superior and very promising technique compared to the conventional methods (see Fig. 1a).

It is particularly important that 3D printing facilitates the fabrication of arbitrary ordered open-cell lattices with high resolution, which is difficult to achieve using traditional techniques. This type of structure has a number of advantages over other geometries, including high porosity (or, alternatively, low relative density), good mechanical strength, extensive pore interconnection and absence of closed cavities, and uniform or well-controlled (e.g. graded) pore distribution. The latter facilitates the controlled distribution of physical properties and pore sizes across the volume, and hence allows to achieve the optimal material and space utilization as well as uniform or gradient material filling / coating. It remains to be seen how por structure, periodicity, volumetric space filling and other factors that can now be 'dialled-in' by 3D printing, allow an accurate assessment of porosity types in battery or supercapacitor electrodes. Porous current collectors or porous active materials more so, are usually described as providing shorter diffusion distance in material and electrolyte for cations, offer distributed electrical wiring throughout the composite, or improve electrolyte soakage among other attributes. If these are universally true to some degree, they can in principle be optimized with CAD designed structures, and the relationship between specific porosity in active or substrate materials on electrochemical behaviour can be properly compared. With the advance of supercapatteries, Li-ion capacitors and other EES system utilizing capacity and capacitance as energy storage processes, the nature of high surface rea materials (all materials of the electrode) is important to be able to quantify.

Several 3D printing methods have been used to fabricate porous metallic materials, which differ by the physical processes involved in the printed layer formation. Most widely used techniques are variations of powder bed fusion (PBF), such as selective laser melting (SLM) and electron beam melting (EBM); material extrusion (ME), for example fused deposition modelling (FDM); vat photopolymerization (VAT-P), e.g. stereolithography (SLA) and digital light processing (DLP) [8, 14, 21, 22, 31-36]. PBF methods are capable of direct fabrication of fully metallic objects. ME and VAT-P techniques usually yield either polymer templates, which are further metallized, or metal–polymer composites that can be conductive and used as prepared (in this case material brittleness issues are quite common). Alternatively, the polymer binder can be removed by post-treatment as in the case of space holder / template-directed techniques discussed earlier in this section.

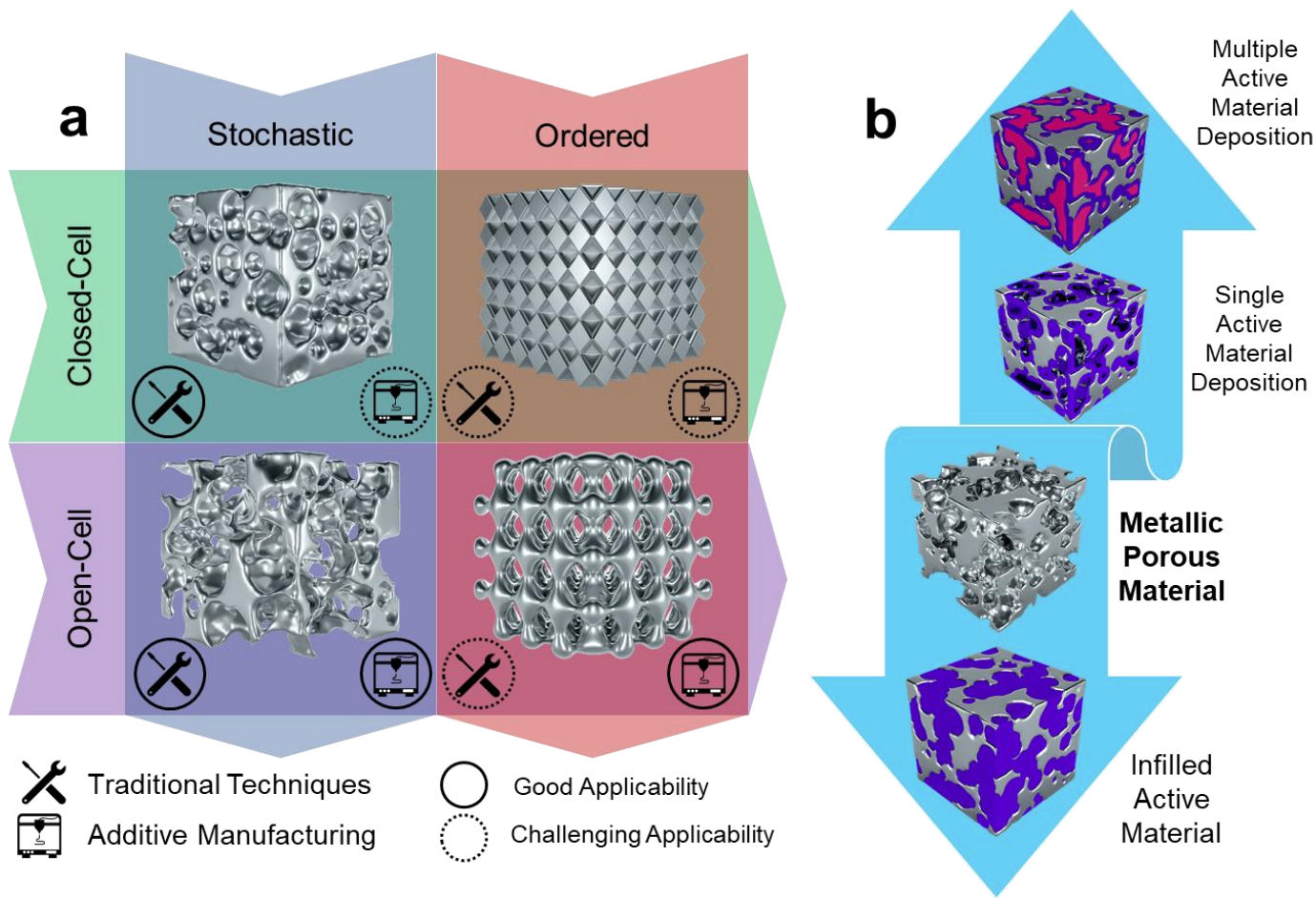

**Figure 1.** (a) Types of metallic porous structures and applicability of traditional and AM techniques to the fabrication of each material type. (b) Main approaches for filling of metallic porous structures with active materials in EES devices such as battery electrodes. We created and rendered these structures using Houdini FX.

Various techniques based on *metallization of templates* (e.g. polymeric foams and lattices) have been reported. These metallization methods include *chemical vapor deposition* (CVD), *physical vapor deposition* (PVD), *electroless plating*, *electrodeposition* and *dip-coating* [12, 22, 23, 33, 37-40]. The template can subsequently be either removed by sintering / dissolution or retained, in which case it can improve mechanical strength and maintain integrity of the structure at the cost of the added weight and volume of usually inactive material. Among the metallization methods, electroless plating is one of the most frequently used due to its simplicity, low cost and relatively uniform resulting metal coating. Also, no specialized expensive equipment is required to perform this technique. Electroless plating involves sensitization and activation of the substrate surface by pre-treatment solutions followed by immersion of substrate into the plating solution containing the desired metal ion [22]. There are several published reviews, which consider the fabrication methods of metallic porous structures in more detail [5, 21, 41].

## 3. Some Properties of Porous Metallic Structures in EES

Structured lattice or porous metals are a class of materials that show a wide variety of interesting properties, and these have been reviewed extensively [2, 3, 26, 31, 34, 42-44]. Aside from modification to ductility, strength, flexibility and effective optical and thermal conductivities caused by interconnected networking of a porous metal, some properties are inherently more interesting to electrochemical energy storage processes. Notably, surface activity, surface chemistry, oxidation (or oxidation resistance), corrosion (resistance), microstructure, and geometry play important active or passive roles in EES devices[14, 34].

As current collectors with high nominal surface area, metals foams such as steel, copper and nickel foams, and other nanostructured metal oxide structures[45], as common examples, are readily used to support active materials in supercapacitor and battery electrodes. Templates metallic periodic porous surface also provide substrates for thin films for batteries, mimicking a porous metal current collector[46]. In practical electrode preparation, metal current collector surfaces are often subject to solvent, water or acid/base exposure, and often involve a thermal treatment in cases where a material is being grown or crystallized on a surface. Often, electrode mass loading is defined by mass, without convincingly uniform coverage of a porous metal surface. This often leaves areas of porous metal that are subject to oxidation, and these oxides can in some cases contribute significantly to the overall electrochemical response[47]. A case in point is work by Geaney et al. who showed [48] how NiO formation on a nickel foam can eventually dominates the response of a Li-ion electrode when the areal mass coverage is low enough to expose oxidized Ni at the porous metal current collector surface.

Figure 2 summarises some of the salient properties or architected or 3D printed metal lattices useful for EES, and by comparison to metal foams (Fig 2b) and solid state synthesis routes, rational design of structure, order and reproducibility is improved. New synthetic approaches have merged with SLA-type 3D printing to produced structure metal lattices on the nanoscale (Fig. 2c). As metal foams tend to be used quite often in supercapacitor and pseudocapacitor research, we recommend that the activity of high surface area (non-noble) oxidizable metal foams be carefully assessed in terms of electrochemical activity in the EES device[50]. The chemistry of Ni, Cu, and other transition metals in acids and bases is now long established[51-54], but when the surface is not entirely covered it will be active in many electrolytes under certain voltage ranges. While this effect is very sensitive in water splitting experiments for a range of metals, it is non-negligible in EES devices where very small mass loadings are used. Adding complexity and surface area to 3D porous metals could exacerbate this issue, unless more chemically stable metals are used or coating methods[55] are improved for lattice metal current collectors or electrodes. Porous metals by their nature, provide spatially inhomogeneous electric fields and locally high current densities, if the lattice features dimension within the porous network are dissimilar. This may pose a problem for EES device electrode coatings that are susceptible to localized corrosion, dissolution, or for electrolyte decomposition in cases of income metal coverage. Additionally, fluctuations in current collector mass can occur when oxidized, or by unexpected (electro)deposition events at exposed regions of a porous or latticed current collector. The often lauded geometrically enhanced activity of nanostructure porous materials[17, 45] for various applications also occurs in other EES devices[56-63], and not always wanted in the case of current collector substrates. On the other hand, modification of engineered porous lattice structures may be beneficial for capacitor-based EES systems, where the actual surface area can be more precisely defined, to provide opportunities to more carefully assess the influence of other parameters (material, voltage range, scan rate, electrolyte etc.) on the capacitance and to decouple intrinsic from geometrical enhancement in such devices[64]. AM grown or 3D printed reproducible porous metal lattices may prove very useful and can be constructed with flat top surfaces to ensure consistent interfacial pressure with the separator and anode, compared to spongy, non-planar metal foams.

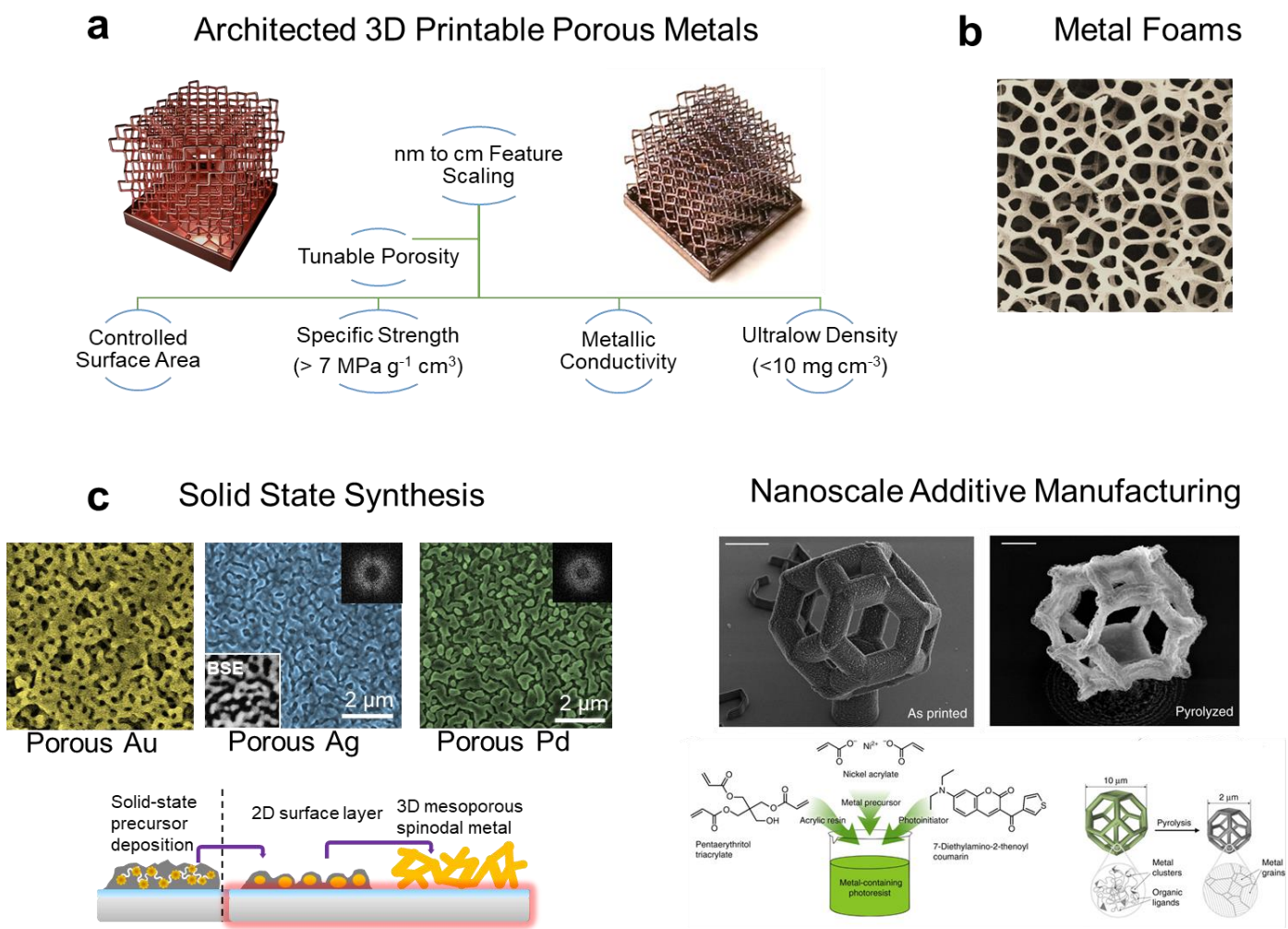

**Figure 2.** (a) Short summary of useful features of additively manufactured, architected porous metals important for EES devices. Other parameters no mentioned are assumed to be controllable by metal type, such as oxidation resistance, oxidation capability, surface reactivity, wettability and flexibility etc. The representative structure was designed and printed by FormLabs 2 stereolithographic 3D printer using the high temperature photoresin. (b) SEM of a typical nickel metal foam structure. (c) Summary of a solid state precursor synthetic route to porous noble metals such as gold and palladium formed as a porous surface film without dealloying. Reproduced from Ref. [49] with permission from the authors. The insets are radially average FFTs of the porous metals confirming constant feature-feature separation in all directions. Nanoscale additive manufacturing using chemical precursors and photo-initiators is capable of producing printed architected metals with nanoscale features. Reproduced from Ref. [8] with permission.

Finally, 3D printing not only allows CAD designed latticed porous metals optimized for a particular electrochemical cell, but can also help reduce the overall mass of current collectors or electrodes. Thin film metallization of 3D printed resin or plastic components requires much less metal than an electrode made entirely from the metal of choice, benefitting costs and overall cell weight. For example, ultralight lattices can be fashioned from 3D printed structures by metal deposition (physical, chemical, or vapor deposition) wit subsequent removal of the underlying polymer or resin template. Designing lattice structure and interconnectivity also ensures well defined electrical conductivity out of plane and in-plane for the electrode and this becomes important for active battery electrode materials that are somewhat less conductive. As case in point in the infilling of lattice metal current collectors to minimize intrinsic resistivity of some materials for thicker electrodes[65]. Overall, adding functional porosity has some usefulness in EES devices, and 3D printing methods can help evolve porous metallic electrode design beyond the limited range and length scales, periodicity, and order of their porosity.

## 4. Conclusions and Outlook

This opinion review looked at the main methods for forming metallic porous structures, foams, and lattices in the context of their using in electrochemical energy storage devices. It is becoming clear that 3D printing and AM approaches in general will become very useful in the future to offset some limitations both in properties and in scaled reproducibility, for porous metals in systems that are inherently very (electro)chemically active. While extension to non-metallic, non-polymeric materials is already possible, rational design of periodicity and porosity in metals with structures and geometries not possible by bulk porosification or physico-chemical assembly in the broadest sense might enable metallic electrode fabrication specifically designed for an EES application that was not possible before. We also hope that the ability to create essentially an lattice or porous metallic structure will allow the community to distinguish intrinsic (surface activity, chemical potential, oxidation state) and geometric (surface area, material fill factor, metallic truss density etc.) as contributors to charge storage in structurally and compositionally very complex electrode formulations.

**Acknowledgements**

This work is supported by Enterprise Ireland Commercialisation Fund as part of the European Regional Development Fund under contract no. CF-2018-0839-P. We also acknowledge funding support from European Union's Horizon 2020 research and innovation programme under grant agreement No 825114, Science Foundation Ireland (SFI) under Award no. 14/IA/2581, and from the Irish Research Council Advanced Laureate Award under grant no. IRCLA/2019/118.

# References

[1] H. Sun, J. Zhu, D. Baumann, L. Peng, Y. Xu, I. Shakir, Y. Huang, X. Duan, Hierarchical 3D electrodes for electrochemical energy storage, Nature Reviews Materials, 4 (2019) 45-60.
[2] J.H. Pikul, J.W. Long, Architected materials for advanced electrochemical systems, MRS Bulletin, 44 (2019) 789-795.
[3] Y. Pang, Y. Cao, Y. Chu, M. Liu, K. Snyder, D. MacKenzie, C. Cao, Additive Manufacturing of Batteries, Advanced Functional Materials, n/a (2019) 1906244.
[4] P.V. Braun, J.B. Cook, Deterministic Design of Chemistry and Mesostructure in Li-Ion Battery Electrodes, ACS Nano, 12 (2018) 3060-3064.
[5] S.J. Yeo, M.J. Oh, P.J. Yoo, Structurally Controlled Cellular Architectures for High-Performance Ultra-Lightweight Materials, Advanced Materials, 31 (2019) 1803670.
[6] A. Sutygina, U. Betke, M. Scheffler, Open-Cell Aluminum Foams by the Sponge Replication Technique, Materials, 12 (2019) 3840.
[7] R. Mines, Theory, Simulation, Analysis and Synthesis for Metallic Microlattice Structures, Metallic Microlattice Structures: Manufacture, Materials and Application, Springer International Publishing, Cham, 2019, pp. 49-65.
[8] A. Vyatskikh, S. Delalande, A. Kudo, X. Zhang, C.M. Portela, J.R. Greer, Additive manufacturing of 3D nano-architected metals, Nature Communications, 9 (2018) 593.
[9] X. Tian, J. Jin, S. Yuan, C.K. Chua, S.B. Tor, K. Zhou, Emerging 3D-Printed Electrochemical Energy Storage Devices: A Critical Review, Adv. Energy Mater., 7 (2017) 1700127.
[10] C. Wang, H. Wang, E. Matios, X. Hu, W. Li, A Chemically Engineered Porous Copper Matrix with Cylindrical Core–Shell Skeleton as a Stable Host for Metallic Sodium Anodes, Advanced Functional Materials, 28 (2018) 1802282.
[11] S. Zekoll, C. Marriner-Edwards, A.K. Ola Hekselman, J. Kasemchainan, C. Kuss, D.E.J. Armstrong, D. Cai, R.J. Wallace, F.H. Richter, J.H.J. Thijssen, P.G. Bruce, Hybrid electrolytes with 3D bicontinuous ordered ceramic and polymer microchannels for all-solid-state batteries, Energy Environ. Sci., 11 (2018) 185 – 201.
[12] H.-W. Zhu, J. Ge, Y.-C. Peng, H.-Y. Zhao, L.-A. Shi, S.-H. Yu, Dip-coating processed sponge-based electrodes for stretchable Zn-MnO2 batteries, Nano Research, 11 (2018) 1554-1562.
[13] T. Song, M. Yan, M. Qian, The enabling role of dealloying in the creation of specific hierarchical porous metal structures—A review, Corrosion Science, 134 (2018) 78-98.
[14] M.S. Saleh, J. Li, J. Park, R. Panat, 3D printed hierarchically-porous microlattice electrode materials for exceptionally high specific capacity and areal capacity lithium ion batteries, Additive Manufacturing, 23 (2018) 70-78.
[15] M. Osiak, H. Geaney, E. Armstrong, C. O'Dwyer, Structuring Materials for Lithium-ion Batteries: Advancements in Nanomaterial Structure, Composition, and Defined Assembly on Cell Performance, J. Mater. Chem. A, 2 (2014) 9433-9460.
[16] A. Vu, Y. Qian, A. Stein, Porous Electrode Materials for Lithium-Ion Batteries – How to Prepare Them and What Makes Them Special, Advanced Energy Materials, 2 (2012) 1056-1085.
[17] J.W. Long, B. Dunn, D.R. Rolison, H.S. White, Three-dimensional battery architectures, Chemical Reviews, 104 (2004) 4463-4492.
[18] O. Al-Ketan, R. Rowshan, R.K. Abu Al-Rub, Topology-mechanical property relationship of 3D printed strut, skeletal, and sheet based periodic metallic cellular materials, Additive Manufacturing, 19 (2018) 167-183.
[19] M.G. Rashed, M. Ashraf, P.J. Hazell, Manufacturing Issues and the Resulting Complexity in Modeling of Additively Manufactured Metallic Microlattices, Applied Mechanics and Materials, 853 (2017) 394-398.
[20] T.A. Schaedler, W.B. Carter, Architected Cellular Materials, Annual Review of Materials Research, 46 (2016) 187-210.
[21] M.G. Rashed, M. Ashraf, R.A.W. Mines, P.J. Hazell, Metallic microlattice materials: A current state of the art on manufacturing, mechanical properties and applications, Materials & Design, 95 (2016) 518-533.
[22] X. Su, X. Li, C.Y.A. Ong, T.S. Herng, Y. Wang, E. Peng, J. Ding, Metallization of 3D Printed Polymers and Their Application as a Fully Functional Water-Splitting System, Advanced Science, 6 (2019) 1801670.
[23] T.A. Schaedler, A.J. Jacobsen, A. Torrents, A.E. Sorensen, J. Lian, J.R. Greer, L. Valdevit, W.B. Carter, Ultralight Metallic Microlattices, Science, 334 (2011) 962.
[24] M.F. Ashby, The properties of foams and lattices, Philosophical Transactions of the Royal Society A: Mathematical, Physical and Engineering Sciences, 364 (2006) 15-30.
[25] M.A. Wettergreen, B.S. Bucklen, B. Starly, E. Yuksel, W. Sun, M.A.K. Liebschner, Creation of a unit block library of architectures for use in assembled scaffold engineering, Computer-Aided Design, 37 (2005) 1141-1149.
[26] S. Singh, N. Bhatnagar, A survey of fabrication and application of metallic foams (1925–2017), Journal of Porous Materials, 25 (2018) 537-554.
[27] Z. Lu, C. Li, J. Han, F. Zhang, P. Liu, H. Wang, Z. Wang, C. Cheng, L. Chen, A. Hirata, T. Fujita, J. Erlebacher, M. Chen, Three-dimensional bicontinuous nanoporous materials by vapor phase dealloying, Nature Communications, 9 (2018) 276.
[28] J. Erlebacher, M.J. Aziz, A. Karma, N. Dimitrov, K. Sieradzki, Evolution of nanoporosity in dealloying, Nature, 410 (2001) 450-453.
[29] Y. Xue, X. Wang, W. Wang, X. Zhong, F. Han, Compressive property of Al-based auxetic lattice structures fabricated by 3-D printing combined with investment casting, Materials Science and Engineering: A, 722 (2018) 255-262.

[30] U. Gulzar, C. Glynn, C. O'Dwyer, Additive manufacturing for energy storage: Methods, designs and materials selection for customizable 3D printed batteries and supercapacitors, arXiv, (2020) 1912.04755.
[31] D. Mahmoud, M.A. Elbestawi, B. Yu, Process–Structure–Property Relationships in Selective Laser Melting of Porosity Graded Gyroids, Journal of Medical Devices, 13 (2019).
[32] S.-I. Park, D.W. Rosen, S.-k. Choi, C.E. Duty, Effective mechanical properties of lattice material fabricated by material extrusion additive manufacturing, Additive Manufacturing, 1-4 (2014) 12-23.
[33] J. Xue, L. Gao, X. Hu, K. Cao, W. Zhou, W. Wang, Y. Lu, Stereolithographic 3D Printing-Based Hierarchically Cellular Lattices for High-Performance Quasi-Solid Supercapacitor, Nano-Micro Lett., 11 (2019) 46.
[34] S.H. Park, M. Kaur, D. Yun, W.S. Kim, Hierarchically Designed Electron Paths in 3D Printed Energy Storage Devices, Langmuir, 34 (2018) 10897 – 10904.
[35] F. Liu, D.Z. Zhang, P. Zhang, M. Zhao, S. Jafar, Mechanical Properties of Optimized Diamond Lattice Structure for Bone Scaffolds Fabricated via Selective Laser Melting, Materials, 11 (2018) 374.
[36] V. Egorov, U. Gulzar, Y. Zhang, S. Breen, C. O'Dwyer, Evolution of 3D Printing Methods and Materials for Electrochemical Energy Storage, arXiv, (2020) 1912.04400.
[37] T. Juarez, A. Schroer, R. Schwaiger, A.M. Hodge, Evaluating sputter deposited metal coatings on 3D printed polymer micro-truss structures, Materials & Design, 140 (2018) 442-450.
[38] J. Song, Y. Chen, K. Cao, Y. Lu, J.H. Xin, X. Tao, Fully Controllable Design and Fabrication of Three-Dimensional Lattice Supercapacitors, ACS Applied Materials & Interfaces, 10 (2018) 39839-39850.
[39] E. Cohen, S. Menkin, M. Lifshits, Y. Kamir, A. Gladkich, G. Kosa, D. Golodnitsky, Novel rechargeable 3D-Microbatteries on 3D-printed-polymer substrates: Feasibility study, Electrochim. Acta, 265 (2018) 690 – 701.
[40] C. Xu, B.M. Gallant, P.U. Wunderlich, T. Lohmann, J.R. Greer, Three-Dimensional Au Microlattices as Positive Electrodes for Li–O2 Batteries, ACS Nano, 9 (2015) 5876-5883.
[41] B. Zhao, A.K. Gain, W. Ding, L. Zhang, X. Li, Y. Fu, A review on metallic porous materials: pore formation, mechanical properties, and their applications, The International Journal of Advanced Manufacturing Technology, 95 (2018) 2641-2659.
[42] X. Ren, R. Das, P. Tran, T.D. Ngo, Y.M. Xie, Auxetic metamaterials and structures: a review, Smart Materials and Structures, 27 (2018) 023001.
[43] X. Yu, J. Zhou, H. Liang, Z. Jiang, L. Wu, Mechanical metamaterials associated with stiffness, rigidity and compressibility: A brief review, Progress in Materials Science, 94 (2018) 114-173.
[44] H. Gu, S. Li, M. Pavier, M.M. Attallah, C. Paraskevoulakos, A. Shterenlikht, Fracture of three-dimensional lattices manufactured by selective laser melting, International Journal of Solids and Structures, 180-181 (2019) 147-159.
[45] P.L. Taberna, S. Mitra, P. Poizot, P. Simon, J.M. Tarascon, High rate capabilities Fe3O4-based Cu nano-architectured electrodes for lithium-ion battery applications, Nature Materials, 5 (2006) 567-573.
[46] H. Zhang, P.V. Braun, Three-Dimensional Metal Scaffold Supported Bicontinuous Silicon Battery Anodes, Nano Lett., 12 (2012) 2778-2783.
[47] S.A. Needham, G.X. Wang, H.K. Liu, Synthesis of NiO nanotubes for use as negative electrodes in lithium ion batteries, Journal of Power Sources, 159 (2006) 254-257.
[48] H. Geaney, D. McNulty, J. O'Connell, J.D. Holmes, C. O'Dwyer, Assessing Charge Contribution from Thermally Treated Ni Foam as Current Collectors for Li-Ion Batteries, Journal of The Electrochemical Society, 163 (2016) A1805-A1811.
[49] C.D. Valenzuela, G.A. Carriedo, M.L. Valenzuela, L. Zúñiga, C. O'Dwyer, Solid State Pathways to Complex Shape Evolution and Tunable Porosity during Metallic Crystal Growth, Sci. Rep., 3 (2013) 2642.
[50] J.C. Park, J. Kim, H. Kwon, H. Song, Gram-Scale Synthesis of Cu2O Nanocubes and Subsequent Oxidation to CuO Hollow Nanostructures for Lithium-Ion Battery Anode Materials, Advanced Materials, 21 (2009) 803-807.
[51] N.L. Peterson, Impurity diffusion in transition-metal oxides, Solid State Ionics, 12 (1984) 201-215.
[52] D. Tuomi, The Forming Process in Nickel Positive Electrodes, J. Electrochem. Soc., 112 (1965) 1-12.
[53] R.D. Armstrong, H. Wang, Behaviour of nickel hydroxide electrodes after prolonged potential float, Electrochimica Acta, 36 (1991) 759-762.
[54] E.E. Kalu, T.T. Nwoga, V. Srinivasan, J.W. Weidner, Cyclic voltammetric studies of the effects of time and temperature on the capacitance of electrochemically deposited nickel hydroxide, Journal of Power Sources, 92 (2001) 163-167.
[55] J. Sudagar, J. Lian, W. Sha, Electroless nickel, alloy, composite and nano coatings – A critical review, Journal of Alloys and Compounds, 571 (2013) 183-204.
[56] D. McNulty, A. Lonergan, S. O'Hanlon, C. O'Dwyer, 3D open-worked inverse opal $TiO_2$ and $GeO_2$ materials for long life, high capacity Li-ion battery anodes, Solid State Ionics, 314 (2018) 195-203.
[57] D. McNulty, E. Carroll, C. O'Dwyer, Rutile TiO2 Inverse Opal Anodes for Li-ion Batteries with Long Cycle Life, High-rate Capability and High Structural Stability, Adv. Energy Mater., 7 (2017) 1602291.
[58] G. Collins, E. Armstrong, D. McNulty, S. O'Hanlon, H. Geaney, C. O'Dwyer, 2D and 3D Photonic Crystal Materials for Photocatalysis and Electrochemical Energy Storage and Conversion, Sci. Technol. Adv. Mater., 17 (2016) 563-582.
[59] E. Armstrong, C. O'Dwyer, Artificial Opal Photonic Crystals and Inverse Opal Structures - Fundamentals and Applications from Optics to Energy Storage, J. Mater. Chem. C, 3 (2015) 6109-6143.
[60] E. Armstrong, D. McNulty, H. Geaney, C. O'Dwyer, Electrodeposited Structurally Stable V2O5 Inverse Opal Networks as High Performance Thin Film Lithium Batteries, ACS Appl. Mater. Interfaces, 7 (2015) 27006-27015.

[61] J.W. Long, B. Dunn, D.R. Rolison, H.S. White, Three dimensional battery architectures., Chem. Rev., 104 (2004) 4463-4492.
[62] J.C. Lytle, Inverse Opal Nanoarchitectures as Lithium-Ion Battery Materials, in: Y. Abu-Lebdeh, I. Davidson (Eds.) Nanotechnology for Lithium-Ion Batteries, Springer Science+Business Media, LLC2013.
[63] N.S. Ergang, J.C. Lytle, K.T. Lee, S.M. Oh, W.H. Smyrl, A. Stein, Photonic crystal structures as a basis for a three-dimensionally interpenetrating electrochemical-cell system, Adv. Mater., 18 (2006) 1750-1753.
[64] Y. Gogotsi, P. Simon, True Performance Metrics in Electrochemical Energy Storage, Science 334 (2011) 917.
[65] Y. Kuang, C. Chen, D. Kirsch, L. Hu, Thick Electrode Batteries: Principles, Opportunities, and Challenges, Advanced Energy Materials, 9 (2019) 1901457.